\documentclass[twocolumn, aps, prl, superscriptaddress]{revtex4}
\usepackage{graphicx}
\usepackage{amssymb}
\usepackage{amsmath}
\def\gsim{\lower -0.3ex \hbox{$>$} \kern -0.75em \lower 0.7ex
\hbox{$\sim$}}
\def\lsim{\lower -0.3ex \hbox{$<$} \kern -0.75em \lower 0.7ex
\hbox{$\sim$}}

\def\Vec#1{{\bf #1}}

\def\vare{\varepsilon}

\begin{document}
%%%%%%%%%%%%%%%%%%%%%%%%%%%%%%%%%%%%%%%%%%%%%%%%%%%%%%%%%%%%%%%%%%%%%%%%%%%%%%%
%
%%%%%%%%%%%%%%%%%%%%%%%%%%%%%%%%%%%%%%%%%%%%%%%%%%%%%%%%%%%%%%%%%%%%%%%%%%%%%%%
\title{
Optical properties of the Hofstadter butterfly
in the Moir\'{e} superlattice
%Optical spectrum of the Hofstadter butterfly
%Optical probing of the Hofstadter butterfly
%Probing Hofstadter butterfly with optical ...
}
%\author{Pilkyung Moon and Mikito Koshino}
%\affiliation{
%School of Computational Sciences, Korea Institute for Advanced Study,
%Seoul, 130--722, Korea,\\
%Department of Physics, Tohoku University, 
%Sendai, 980--8578, Japan}
\author{Pilkyung Moon}
\affiliation{
School of Computational Sciences, Korea Institute for Advanced Study,
Seoul, 130--722, Korea}
\author{Mikito Koshino}
\affiliation{
Department of Physics, Tohoku University, 
Sendai, 980--8578, Japan}
\date{\today}

\begin{abstract}
We investigate the optical absorption spectrum and
the selection rule for the Hofstadter butterfly
in twisted bilayer graphene under magnetic fields.
We demonstrate that the absorption spectrum
exhibits a self-similar recursive pattern
reflecting the fractal nature of the energy spectrum.
We find that the optical selection rule 
has a nested self-similar structure as well, 
and it is governed by the conservation
of the total angular momentum summed over different hierarchies.
\end{abstract}
\maketitle

\begin{figure*}
\begin{center}
%\leavevmode\includegraphics[width=0.9\hsize]{Fig_2.65.eps}
\leavevmode\includegraphics[width=0.9\hsize]{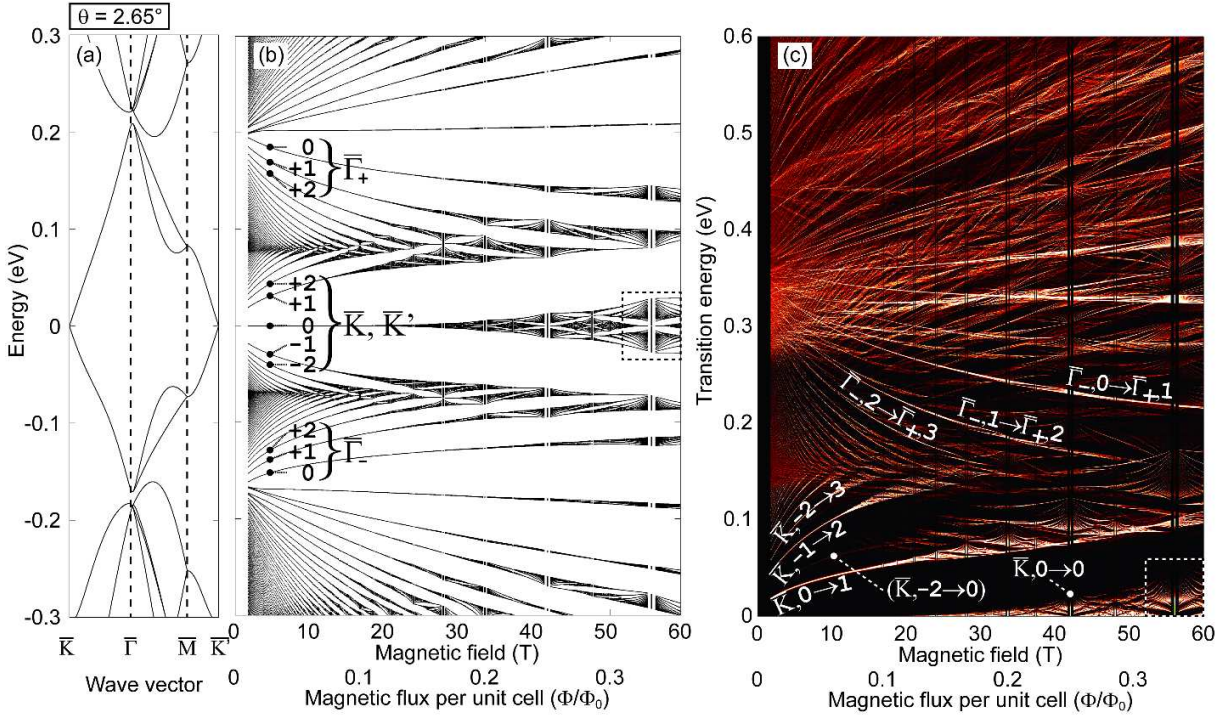}
\end{center}
\caption{
(a) Band structure at $B=0\,\mathrm{T}$ 
and (b) energy spectrum as a function of $B$,
calculated for TBG with $\theta = 2.65^\circ$.
(c) Intensity map of the optical conductivity
for right circularly polarized light
(Re $\sigma_{\mathrm{+}}$) of the same TBG.
}
\label{fig_2_65}
\end{figure*}

\begin{figure}
\begin{center}
\leavevmode\includegraphics[width=0.95\hsize]{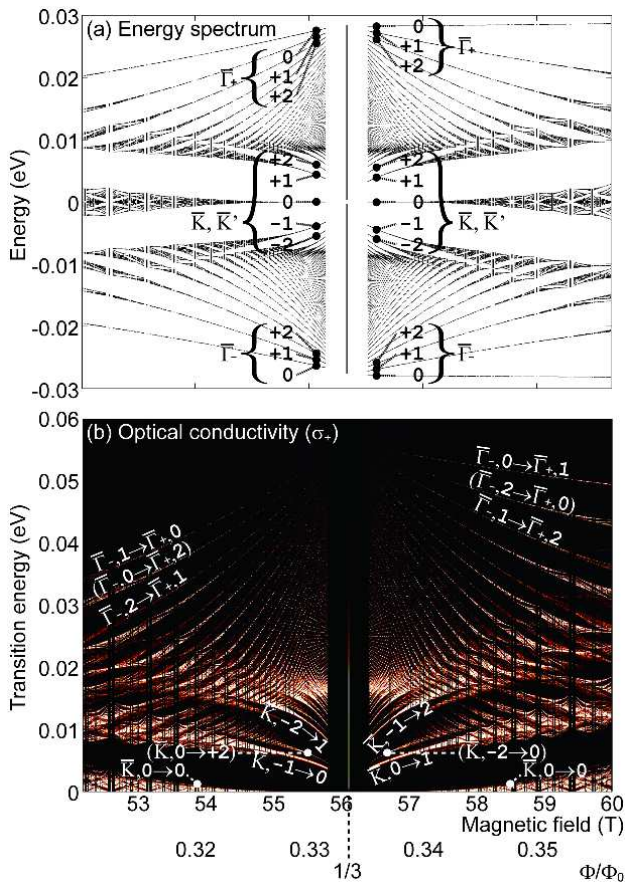}
\end{center}
\caption{
(a) Energy spectrum and (b) optical conductivity for 
right circularly polarized light (Re $\sigma_{\mathrm{+}}$) 
of TBG with $\theta = 2.65^\circ$,
near the $n = 0$ Landau level 
around $\Phi/\Phi_0 = 1/3$
[the regions marked by dashed boxes in Figs.\ \ref{fig_2_65}(b) and (c)].
}
\label{fig_2_65_subgeneration}
\end{figure}

\begin{figure*}
\begin{center}
\leavevmode\includegraphics[width=0.9\hsize]{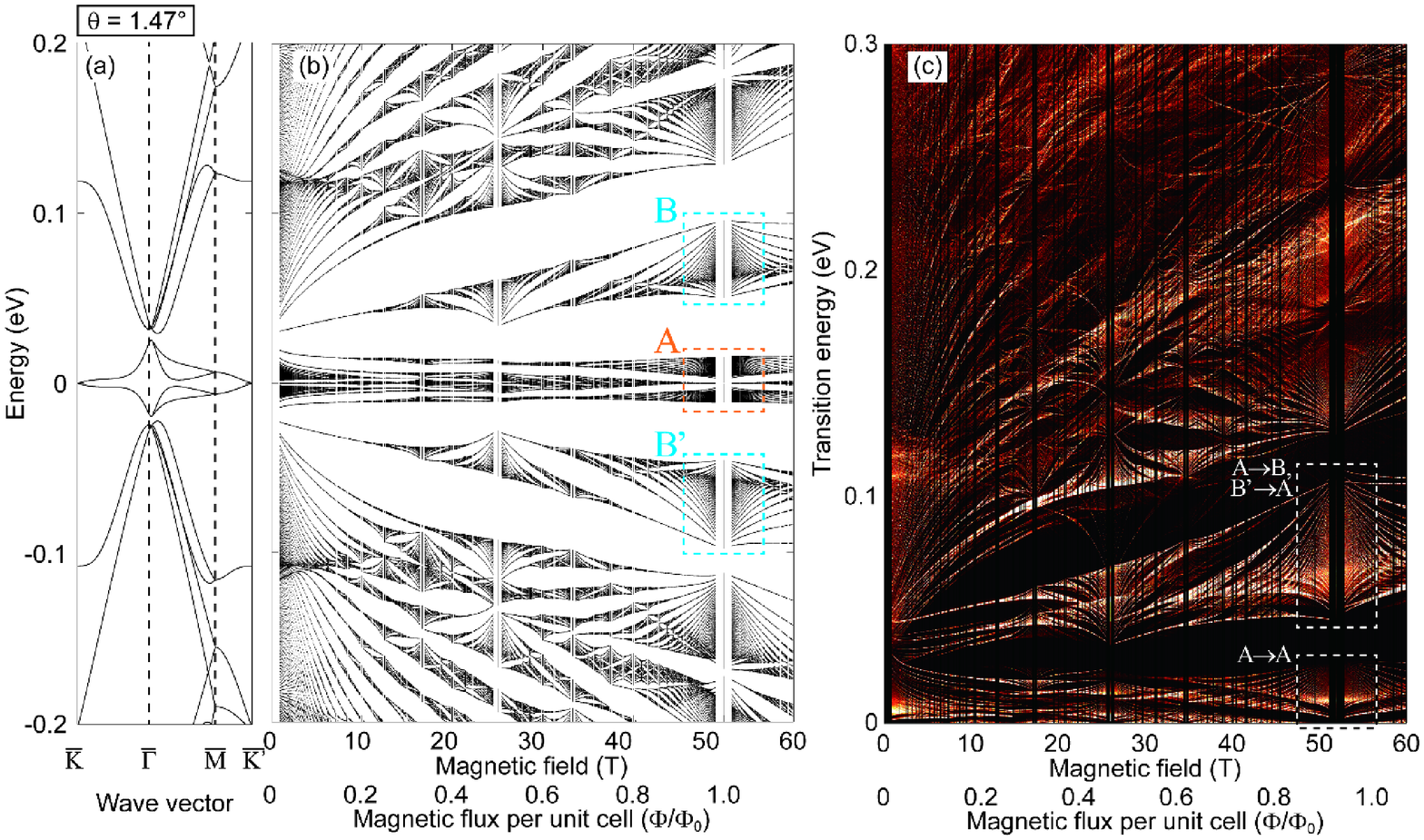}
\end{center}
\caption{
(a) Band structure at $B=0\,\mathrm{T}$ 
and (b) energy spectrum as a function of $B$,
calculated for TBG with $\theta = 1.47^\circ$.
(c) Intensity map of the optical conductivity
(Re $\sigma_{\mathrm{+}}$) of the same TBG.
}
\label{fig_1_47_energy_spectrum}
\end{figure*}

%%%%%%%%%%%%%%%%%%%%%%%%%%%%%%%%%%%%%%%%%%%%%%%%%%%%%%%%%%%%%%%%%%%%%%%%%%%%%%%
%
%%%%%%%%%%%%%%%%%%%%%%%%%%%%%%%%%%%%%%%%%%%%%%%%%%%%%%%%%%%%%%%%%%%%%%%%%%%%%%%

% Introduction: Hofstadter butterfly
Electrons under the simultaneous influence of
a periodic potential and a magnetic field
exhibit a self-similar energy spectrum
due to the competition between Bragg reflection and Landau quantization
\cite{zak1964magnetic,langbein1969tight,hofstadter1976energy,
schlosser1996internal,albrecht2001evidence,geisler2004detection}.
Such a fractal band structure,
which is called the Hofstadter butterfly,
appears when the magnetic flux $\Phi$
per a unit cell becomes as large as $\Phi_0 = h/e$.
%To realize this effect in an experimetally feasible magnetic field,
%a modulation period needs to be much longer than the atomic scale,
%and the first experimental attempts were made to find this effect
%in artificial superlattices
%\cite{schlosser1996internal,albrecht2001evidence,geisler2004detection},
Recently, it was theoretically proposed
\cite{bistritzer2011moireprb,PhysRevB.85.195458,wang2012fractal}
that the Hofstadter spectrum emerges 
in twisted bilayer graphene (TBG)
\cite{lopes2007graphene,hass2008multilayer,trambly2010localization,shallcross2010electronic,morell2010flat}
in moderate magnetic field,
owing to a long-period Moir\'{e} pattern
in the misoriented lattice structure
\cite{hermann2012periodic}.
The evidence of the fractal nature was observed 
in the recent magnetotransport measurements
in Moir\'{e} superlattice
composed of graphene and hexagonal boron nitride
\cite{dean2013hofstadter,ponomarenko2013cloning,hunt2013massive}.

The realization of fractal energy band
leads to an interest in how the optical property looks like
in this unique spectrum.
%which enables a realistic access
%to the mystic butterflylike gaps.
The magneto-optical absorption measurement has been widely used
to investigate the Landau level structure of graphene systems
\cite{sadowski2006landau,jiang2007infrared,deacon2007cyclotron,plochocka2008high,gusynin2006transport,PhysRevB.75.155430,koshino2008magneto,crassee2010giant}.
For TBG, the optical spectrum at zero magnetic field 
was recently studied theoretically
\cite{wang2010stacking, chen2011stacking, tabert2013optical,
PhysRevB.87.205404,stauber2013optical_arxiv},
and experimentally \cite{wang2010stacking,ohta2012evidence},
while the magneto-optical property remains to be
explored.
In the literature, the optical absorption
of the Hofstadter butterfly was studied
for a simple square lattice \cite{vogl2002self},
while a detailed study is needed
to specify the optical selection rule in the recursive spectrum.

The purpose of this paper is to reveal
the optical absorption of the Hofstadter butterfly in TBG.
TBG offers an excellent platform to investigate
the fractal spectrum in optics, because
the energy scale of the fractal structure 
can be relatively large due to the direct coupling of 
the two graphene bands in the low-energy region.
Also, the rich variety of the 
spectrum sensitively depending on the rotation angle
is a fascinating subject to investigate.
Here we calculate the dynamical conductivity of TBG
as a function of magnetic field, and demonstrate
that the optical spectrum
exhibits the hierarchical recursive pattern.
We also find that the selection rules for optically allowed 
transitions exhibit a nested self-similar structure,
which is governed by the conservation of the angular momentum
summed over different hierarchies.

% Theory: B!=0T
In the following, we investigate TBGs with 
$\theta = 2.65^\circ$ and $1.47^\circ$,
where $\theta$ is the relative rotation angle 
between two graphene layers.
The period of the Moir\'{e} superlattice
is given by $L_{\rm M} = a/[2\sin(\theta/2)]$
($L_{\rm M}$ = 5.33\,nm and 9.59\,nm for $\theta = 2.65^\circ$ and $1.47^\circ$, respectively)
where $a$ is the lattice constant of graphene
\cite{shallcross2010electronic,hermann2012periodic,PhysRevB.87.205404}.
In TBG, the graphene's linear band is 
folded by the periodic interlayer
coupling, giving the energy scale of 
$\varepsilon_{\rm M} \sim \pi \hbar v / L_{\rm M}$
with the linear band velocity $v$.
In small rotation angles less than $2^\circ$,
in particular, the energy band is strongly modified
\cite{trambly2010localization,morell2010flat,luican2011single,bistritzer2011moirepnas,macdonald2011materials,PhysRevB.85.195458},
because $\varepsilon_{\rm M}$ becomes comparable to 
the magnitude of the interlayer coupling.
The two rotation angles
considered in this paper exhibit notably different band
structures, where $\theta=2.65^\circ$ still keeps monolayerlike dispersion
near the Dirac point, while $\theta=1.47^\circ$ exhibits
an almost flat band at the Dirac point.

We model %calculate the eigen wavefunctions of 
TBG with a single-orbital
(carbon $2p_z$ orbital)
tight-binding Hamiltonian, 
\begin{eqnarray}
&& H_{\rm TBG} = -\sum_{\langle i,j\rangle}
t(\Vec{R}_i,\Vec{R}_j) e^{i\phi_{ij}}
|\Vec{R}_i\rangle\langle\Vec{R}_j| + {\rm H.c.},
\end{eqnarray}
where $\Vec{R}_i$ and $|\Vec{R}_i\rangle$ 
represent the lattice point and the atomic state at site $i$, respectively,
$t(\Vec{R}_i - \Vec{R}_j)$ is
the transfer integral between
%the sites $i$ and $j$,
site $i$ and site $j$.
The phase factor 
$\phi_{ij} = -(e/\hbar)\int_j^i \Vec{A}(\Vec{r})\cdot d\Vec{r}$
is a Peierls phase where  
$\Vec{A}(\Vec{r}) = (0,Bx,0)$ is the vector potential
giving a uniform magnetic field $B$ perpendicular to the layers.
For $t(\Vec{R})$, we adopt an approximation
\cite{slater1954simplified,PhysRevB.85.195458,trambly2010localization,PhysRevB.87.205404},
\begin{eqnarray}
 && -t(\Vec{d}) = 
V_{pp\pi}(d)\left[1-\left(\frac{\Vec{d}\cdot\Vec{e}_z}{d}\right)^2\right]
+ V_{pp\sigma}(d)\left(\frac{\Vec{d}\cdot\Vec{e}_z}{d}\right)^2,
\nonumber \\
&& V_{pp\pi}(d) =  V_{pp\pi}^0 % -\tilde\gamma_0
e^{- \frac{d-a_0}{r_0}},\quad
V_{pp\sigma}(d) =  V_{pp\sigma}^0 %\tilde\gamma_1
e^{- \frac{d-d_0}{r_0}},
\label{eq_transfer_integral}
\end{eqnarray}
where
$V_{pp\pi}^0 \approx -2.7\,\mathrm{eV}$ 
is the transfer integral between 
the nearest-neighbor atoms of monolayer graphene,
and $V_{pp\sigma}^0 \approx 0.48\,\mathrm{eV}$ is that
between vertically located atoms on the neighboring layers.
$r_0$ is the decay length of the transfer integral,
and is chosen as $ 0.045\,\mathrm{nm}$ so that 
the next nearest intralayer coupling becomes $0.1 V_{pp\pi}^0$
\cite{deacon2007cyclotron}.
$a_0 \approx 0.142$ nm
is the distance between the nearest carbon atoms in graphene,
and $d_0 \approx 0.335\,\mathrm{nm}$ is the interlayer spacing.

The other three orbitals of carbon ($2s$ orbital and two $2p$ orbitals)
are safely ignored, since they form $sp_2$ hybridized orbitals
of which energies are far from Dirac point.
The single-orbital tight-binding method has been widely used to
investigate electronic properties of carbon-based materials
(e.g., graphite, carbon nanotube, and graphene ) 
\cite{dresselhaus2002intercalation,
ando2005theory,neto2009electronic,peres2010colloquium},
and the approximate values of the parameters are well established
\cite{dresselhaus2002intercalation}.

For the basis to construct the Hamiltonian matrix, 
we only take the wavefunctions of 
low-lying Landau levels of monolayer graphene in 
$|\vare| \lsim 1.5\,\mathrm{eV}$, 
and compose the Hamiltonian matrix by writing 
$H_{\rm TBG}$ in terms of the reduced basis
\cite{PhysRevB.85.195458}.
% Theory: Kubo formula
The dynamical conductivities
for right ($\sigma_+$)
and left ($\sigma_-$)
circularly polarized light
can be obtained by \cite{inoue1962landau,toy1977minority}
\begin{eqnarray}
\sigma_{\pm}(\omega) &=& 
\frac{e^2\hbar}{i S}
\sum_{\alpha,\beta}
\frac{f(\varepsilon_\alpha)-f(\varepsilon_\beta)}
{\varepsilon_\alpha-\varepsilon_\beta}
\frac{|\langle\alpha|v_{\pm}|\beta\rangle|^2}{\varepsilon_\alpha-\varepsilon_\beta+\hbar\omega+i\eta},
\label{eq_dynamical_conductivity}
\end{eqnarray}
where $S$ is the area of the system, 
$f(\varepsilon)$ is the Fermi distribution function,
$\varepsilon_{\alpha}$ ($\varepsilon_{\beta}$) and
$|\alpha\rangle$ ($|\beta\rangle$)
represent the eigenenergy and the eigenstate of the system,
$v_{\pm} = (v_x \pm i v_y)/\sqrt{2}$,
$v_x=-(i/\hbar)[x,H]$ is the velocity operator, and 
$\eta$ is a phenomenological broadening
which is taken as $\eta = 0.05\,\mathrm{meV}$
for $\theta = 2.65^\circ$
and $0.025\,\mathrm{meV}$ for $1.47^\circ$.
The optical absorption intensity
%of light incident perpendicular to
of incident light perpendicular to a two-dimensional system,
at photon energies $\hbar\omega$,
is approximately given by
$(4\pi/c){\rm Re}\,\sigma_{\pm}(\omega)$.
%$(4\pi/c){\rm Re}\,\sigma_{xx}(\omega)$. 
We fix the temperature to 0
and assume the Fermi energy at the charge neutrality point.

% Band structure (B=0T)
Figure \ref{fig_2_65}(a) shows the band structure
of TBG with $\theta=2.65^\circ$ at zero magnetic field.
$\bar{K}$, $\bar{K}'$, $\bar{M}$, and $\bar{\Gamma}$
represent the high-symmetry points of
the reduced Brillouin zone of TBG.
%calculated by the tight-binding model, Eq.\ ???.
The lowest conduction and valence bands, 
lying in the energy range from $-0.2$ eV to 0.2 eV,
are characterized by the monolayerlike linear dispersion
near $\bar{K}$ and $\bar{K}'$,
saddle point at $\bar{M}$,
and a holelike (electronlike) pocket
in the conduction (valence) band at $\bar{\Gamma}$
\cite{PhysRevB.85.195458,PhysRevB.87.205404}.

% Band structure (B!=0T)
In Fig.\ \ref{fig_2_65}(b), we plot
the energy spectrum against the magnetic field $B$ for the same TBG.
In weak magnetic field,
the lowest conduction band is composed of
monolayerlike Landau levels at $\bar{K}$ and 
$\bar{K}'$ which behave in proportion to $\sqrt{B}$,
and also the holelike Landau levels at $\bar{\Gamma}$
which move downward in increasing $B$
\cite{PhysRevB.85.195458}.
We label the Landau levels at $\bar{K},\bar{K}'$
by $(\bar{K},n)$ and $(\bar{K}',n)$ with
 $n = 0, \pm 1, \pm 2, \cdots$, 
and those associated with $\bar{\Gamma}$
by $(\bar{\Gamma}_+,n)$ and $(\bar{\Gamma}_-,n)$ with $n = 0, 1, 2\cdots$,
for the positive and the negative energy parts, respectively.
The labeling of the Landau levels is indicated in Fig.\ \ref{fig_2_65}(b).

% Magneto-optical conductivity, theta=2.65deg.
Figure \ref{fig_2_65}(c) shows
a density plot of the optical absorption intensity 
for circularly polarized light (${\rm Re}\, \sigma_{\mathrm{+}}$)
on the space of the transition energy $\hbar \omega$
and the magnetic field $B$,
calculated for TBG with $\theta=2.65^\circ$.
The spectrum in the weak field regime $(\Phi/\Phi_0 < 0.1)$
consists of sharp peaks associated with the transitions
between discrete Landau levels.
Below the transition energy of $\hbar\omega \approx 0.15\,\mathrm{eV}$,
which is the energy span between the van Hove singularities
in Fig.\ \ref{fig_2_65}(a),
we observe the monolayerlike absorption peaks 
corresponding to the transitions from
$(\bar{K},-n)$ to $(\bar{K},n+1)$.
Above $0.15\,\mathrm{eV}$, we can see the series of spectral lines
which go down with increasing $B$,
and they are associated with
the transitions from $(\bar{\Gamma}_-, n)$
to $(\bar{\Gamma}_+,n+1)$.
The transitions between the different families of Landau levels,
such as $\bar{K} \to \bar{\Gamma}_+$, are negligibly small.

% selection rule
The selection rule for the optically allowed transition
is expressed as $|n| \rightarrow |n|+1$
in either transition series.
This is because the relative angular momentum of a state
in the Landau level $n_2$ to one in $n_1$
with the same center coordinate is $(|n_2|-|n_1|)\hbar$
regardless of the electronlike or holelike level,
and a right circularly polarized photon carries an angular momentum
of $+\hbar$.
For the left circularly polarized light  ($\sigma_{\mathrm{-}}$),
the photon's angular momentum becomes
$-\hbar$
and the selection rule changes to $|n| \rightarrow |n|-1$.
In TBG, the absorption peak positions generally differ 
between $\sigma_{\mathrm{+}}$ and $\sigma_{\mathrm{-}}$ (not shown)
because of the electron-hole asymmetry in the energy spectrum
\cite{luican2011single,PhysRevB.87.205404}.
The system should exhibit a significant circular dichroism
when the frequency hits a peak
of $\sigma_{\mathrm{+}}$ or $\sigma_{\mathrm{-}}$.
% trigonal warping etc.
Besides the major peaks
described by the above selection rule,
we also observe additional, minor peaks
corresponding to the transition
$|n| \rightarrow |n|+3m+1$ ($m$: integer).
This is a direct consequence of the trigonal warping of
zero-field band structure
which hybridizes $|n|$ and $ |n|+3$
\cite{PhysRevB.75.155430,morimoto2012faraday}.
%In the $\bar{K}$ and $\bar{K}'$ levels,
%the intensity of subordinate peaks are
%much smaller than that of the major peaks,
%while in the $\bar{\Gamma}$ levels,
%they have comparable intensity to the major peaks,
%in accordance with the strong trigonal warping in
%the $\bar{\Gamma}$ pocket. \cite{PhysRevB.85.195458}
The optical spectrum above $\hbar \omega \sim 0.3\,\mathrm{eV}$
is complicated because it includes the absorption peaks 
associated with the Landau levels in the higher energy band.

% Strong field regime
In increasing magnetic field,
as we can see from Fig.\ \ref{fig_2_65}(b),
the spectrum evolves into Hofstadter butterfly,
where each Landau level splits into
sub-generation levels with a smaller energy scale.
The optical spectrum in Fig.\ \ref{fig_2_65}(c)
also evolves into a fractal structure in this regime,
reflecting the hierarchy of the recursive energy structures.
The spectrum is characterized by 
$\Phi/\Phi_0$, i.e., the number of flux quanta
penetrating the Moir\'{e} unit cell,
where each Landau level forms $2p$ subbands
when $\Phi/\Phi_0$ is a rational number $p/q$ 
($p$ and $q$ are coprime integers).
Here the factor 2 in $2p$ comes from the number of layers.

Figure \ref{fig_2_65_subgeneration}
shows the energy and optical spectra near the $n = 0$ Landau level 
around $\Phi/\Phi_0 = 1/3$, corresponding to 
the regions enclosed by
the dashed boxes in Figs.\ \ref{fig_2_65}(b) and (c).
%The spectrum is missing in the close vicinity of $\Phi/\Phi_0 = 1/3$
%due to the limitation of numerical calculation for the fractions 
%with large demoninator $q$.
At $\Phi/\Phi_0 = 1/3$, the $n = 0$ Landau level 
becomes a pair of energy bands touching at the 
zero energy, and 
the energy dispersion in the magnetic Brillouin zone
is shown to have a similar structure to the lowest energy band at $B=0\,\mathrm{T}$.
When the flux is shifted from 1/3, 
the band splits into the second-generation Landau levels,
which can be labeled by $(\bar{K},n')$ and $(\bar{\Gamma}_\pm,n')$
as shown in Fig.\ \ref{fig_2_65_subgeneration}(a),
similarly to the first-generation Landau levels.
Correspondingly, the optical spectrum in
$\Phi/\Phi_0 > 1/3$
in Fig.\ \ref{fig_2_65_subgeneration}(b)
resembles that for the first-generation in
$\Phi/\Phi_0 > 0$,
and the selection rule becomes $|n'| \rightarrow |n'|+1$.
We also see the additional excitations $|n'| \rightarrow |n'|+1+3m$
due to the trigonal warping.

It should be noted that these transitions occur
inside the same parent Landau level $n=0$,
which were originally forbidden in the weak magnetic field 
$\Phi/\Phi_0 < 0.1$.
They come to be allowed in the butterfly regime, 
because the second-generation Landau levels have the 
additional angular momentum depending on $n'$, 
and can absorb the 
photon's angular momentum. In
$\Phi/\Phi_0 < 1/3$,
the selection rule becomes  $|n'| \rightarrow |n'|-1$,
since the residual magnetic field from 
$\Phi/\Phi_0 = 1/3$ is negative, and 
the angular momenta for the second-generation
Landau levels reverse 
the sign.
The same structure can be seen also at
$\Phi/\Phi_0 = 1/q$ for every integer $q$.

% different parant generation
We also have the transitions between
the sub-Landau levels belonging to 
different parent Landau levels.
The global address for an energy level near $\Phi/\Phi_0 = 1/q$
is written like $(n; \bar{K},n')$ or $(n; \bar{\Gamma}_+,n')$, etc.,
where $n$ and $n'$ are the Landau level indices 
for the first generation and the second generation,
respectively.
%While the monolayer's Landau level structure
%loses its original form in the butterfly regime, 
%we can determine the index 
%by counting the number of states
%from the charge neutrality point.
The transition occurs only inside the same valley
$\bar{K}$, $\bar{K}'$ and $\bar{\Gamma}$ 
($\bar{\Gamma}_\pm \to \bar{\Gamma}_\pm$ and 
$\bar\Gamma_\pm \to \bar\Gamma_\mp$ are both allowed).
In $\Phi/\Phi_0 \geq 1/q$, the selection rule is found to be
$|n_i| + |n'_i| + 1 = |n_f| + |n'_f|$ (in modulo 3), 
where $i$ and $f$ are for the initial and final states, respectively,
indicating that the total angular momentum 
is written as the sum over the first and the second generations.
In $\Phi/\Phi_0 \leq 1/q$, the selection rule changes to
$|n_i| - |n'_i| + 1 = |n_f| - |n'_f|$ (in modulo 3),
because the sign of the angular momentum in the
second-generation level becomes opposite.

% experimental condition
The condition to observe the sub-generation Landau levels
can be estimated by comparing the width of Moir\'{e}-modulated
Landau level ($w$) and the disorder broadening ($\Gamma$). 
For zero-th Landau level, for example, the order of $w$ is estimated as
$w \sim V^0_{pp\sigma} e^{-\pi/(3\sqrt{3}\Phi)}$,
when the inter Landau level mixing is not very strong.
When $\Phi$ exceeds 0.1, $w$ rapidly grows 
from an exponentially small value to a significant magnitude.
In the long-range disorder potential, 
$\Gamma$ is shown to be proportional to $\sqrt{B}$
\cite{shon1998quantum}, 
so that the condition $w > \Gamma$ is
more easily achieved in the higher magnetic field.  
In short-range scatterers such as vacancies, on the other hand, 
the low-lying Landau levels are strongly broadened 
by the impurity levels at zero energy
\cite{peres2006electronic}, 
and it may mask the butterfly structure near the zero-th Landau level.

%1.47
Figure \ref{fig_1_47_energy_spectrum} shows 
the energy spectrum and the optical absorption spectrum
of TBG with a smaller rotation angle $\theta = 1.47^\circ$.
The lowest energy band shrinks drastically near the Dirac point
in this case \cite{trambly2010localization},
and the monolayerlike absorption peaks are almost invisible.
Instead, the fractal spectrum is clearly seen
in smaller magnetic field than in  $\theta = 2.65^\circ$,
due to the larger Moir\'{e} superlattice period.
At the magnetic flux $\Phi/\Phi_0 = 1$,
the regions $A$, $B$ and $B'$ 
[enclosed by dashed boxes in Fig.\ \ref{fig_1_47_energy_spectrum}(b)]
belong to the original monolayer's Landau levels 
of $n=0$,1 and $-1$, respectively.
In the optical absorption spectrum, Fig.\ \ref{fig_1_47_energy_spectrum}(c), 
the low-energy peaks are attributed to the transitions between these blocks,
and the optical transition between second-generation Landau levels
is governed by the selection rule discussed above.

% Conclusion
We have theoretically investigated
the optical absorption spectra of
Hofstadter butterfly in TBG under magnetic field.
In weak magnetic field, the absorption spectrum
consists of sharp peaks
associated with the transitions
between discrete Landau levels.
In increasing magnetic field,
the spectrum gradually
evolves
into a fractal structure,
reflecting the hierarchy of the recursive
energy spectrum.
We have shown that the optical selection rule
exhibits a nested self-similar structure,
which is governed by the conservation
of the angular momentum
summed over different hierarchies.

% Acknowledgement
Computational resources have been provided by CAC of KIAS (P.M.).
This project has been funded by JSPS Grant-in-Aid for
Scientific Research No.\ 24740193 and
No. 25107005 (M.K.).

%\bibliography{Qiqqa2BibTexExport}

\bibliography{Qiqqa2BibTexExport,koshino}

%\begin{thebibliography}{99}
%\bibitem{comment_intralayer}
%In this paper, we numerically calculated
%both the interlayer and intralayer
%interactions.
%The energy scale of the resulting spectrum
%is slightly different from 
%that in Ref.\ \cite{PhysRevB.85.195458},
%which is obtained by using an analytic expression
%for intralayer elements.
%However, the overall aspect of the spectrum agrees well.
%\end{thebibliography}

\end{document}